\def\ArtDir{main/}

\documentclass{ws-ijgmmp}
\def\ArtDir{main/}
\usepackage{cite}
\usepackage{comment}

\begin{document}

\markboth{Mar\'ia-Jos\'e Guzm\'an and Shymaa Khaled Ibraheem}{ Classification of primary constraints for new general relativity in the premetric approach }

\catchline{}{}{}{}{}

\title{Classification of primary constraints for new general relativity\\ in the premetric approach}

\author{Mar\'ia-Jos\'e Guzm\'an}

\address{Departamento de F\'isica y Astronom\'ia, Facultad de Ciencias,\\ Universidad de La Serena, Av. Juan Cisternas 1200, 1720236 La Serena, Chile,
\email{maria.j.guzman.m@gmail.com}}

\author{Shymaa Khaled Ibraheem}

\address{Mathematics Department, Faculty of Science,
Ain Shams University,11566 Cairo, Egypt\\
Centre for Theoretical Physics, British University in Egypt, El Sherouk City 11837, Egypt.
shymaa.khaled90@gmail.com}

\maketitle


\newcommand{\cc}{\cite}

\begin{abstract}
We introduce a novel procedure for studying the Hamiltonian formalism of new general relativity (NGR) based on the mathematical properties encoded in the constitutive tensor defined by the premetric approach. We derive the canonical momenta conjugate to the tetrad field and study the eigenvalues of the Hessian tensor, which is mapped to a Hessian matrix with the help of indexation formulas. The properties of the Hessian matrix heavily rely on the possible values of the free coefficients $c_i, i=1,2,3$ appearing in the NGR Lagrangian. We find four null eigenvalues associated with trivial primary constraints in the temporal part of the momenta. The remaining eigenvalues are grouped in four sets, which have multiplicity 3, 1, 5 and 3, and can be set to zero depending on different choices of the coefficients $c_i$. There are nine possible different cases when one, two, or three sets of eigenvalues are imposed to vanish simultaneously. All cases lead to a different number of primary constraints, which are consistent with previous work on the Hamiltonian analysis of NGR by Blixt et al. (2018).
\end{abstract}


\section{Introduction}

General relativity extensions based on modifications to the geometric foundations of gravity are nowadays an active field of research, which are motivated by its success at producing new physics in the cosmological arena. This could help to alleviate the well-known paradigms appearing in modern cosmology such as early and late accelerated phases of the Universe. Moreover, due to their inherent gauge structure, this kind of theories are promising for giving clues to the study of canonical quantum gravity.  For a long time, it has been known two alternative descriptions of the  gravitational phenomena, which change the underlying geometric description of gravity based on curvature. They are known as the teleparallel equivalent of general relativity (TEGR) and the symmetric teleparallel gravity (STEGR). The torsion and non-metricity of spacetime are the geometrical entities that encode gravitation in TEGR and STEGR, respectively \cc{BeltranJimenez:2019tjy}. 

One particular modified teleparallel gravity that we are interested to study is based in a manifold equipped with a teleparallel connection, and it is generically known as new general relativity (NGR) \cc{hayashi1979new}. It consists of a generalization of the TEGR, where the Lagrangian is built from a linear combination of three parity-even scalars. These terms are quadratic forms of the torsion tensor, therefore quadratic on first-order derivatives of the tetrad field. They behave as scalars under local Lorentz transformations and as a density under general changes of coordinates \cc{pellegrini1963}. Ever since NGR was formulated, several topics have been studied within this theory: spherically symmetric geometries have been found \cc{Fukui:1981si,Kawai:1989qt}, cosmological applications have been reported \cc{mikhial1995cosmological}, and recently polarization modes of gravitational waves in the linearized theory have been studied \cc{hohmann2018propagation}. In recent years, NGR has been used as a starting point for building more general teleparallel theories of gravity. Although this theory modifies GR in a nontrivial way, there are certain indications that NGR would not be a healthy theory unless it reduces to the GR case \cc{jimenez2020non}. To confirm these claims with additional investigations is pending, however, it is anyway interesting to study NGR on itself as a toy model that is being actively used as a  building block for modified teleparallel gravities, for example,  $f(T_{ax},T_{vec},T_{ten})$ theories \cc{Bahamonde:2017wwk} or teleparallel Horndeski gravity \cc{Bahamonde:2019shr}, among others.

A crucial point to study in any physical theory is its predictability and internal consistency, in particular to prove that it contains a well-defined number of degrees of freedom (d.o.f.) and the Cauchy problem is well-posed. The best path in this direction seems to be the Hamiltonian formalism delineated by the Dirac-Bergmann algorithm for constrained systems. This robust method clearly exhibits the physical content of the theory through the determination of first- and second-class constraints, where the former can sometimes be associated with gauge transformations of the theory, and the later can be organized as pairs of spurious conjugated variables, giving as a result a well defined and non ambiguous number of d.o.f. \cite{anderson1951constraints,dirac1964lectures,sundermeyer1982constrained,henneaux1994quantization,sundermeyer2014symmetries}. 

The Hamiltonian analysis of NGR has been partially studied in Refs. \cc{blixt2019ahamiltonian,blixt2019bhamiltonian,blixt2019gauge,Hohmann:2019sys,mitric2019canonical}, where we highlight the analysis that has been performed in \cc{blixt2019ahamiltonian}. In this work, the authors have found the canonical  Hamiltonian and primary constraints for all possible different cases allowed by the three free parameters of the theory. However, the full calculation of constraint algebra is still missing and implies a huge effort, to which we would like to contribute to alleviating. For this, it is helpful to consider previous research on the Hamiltonian formalism for TEGR and weight the tools that have been used in the past. The Hamiltonian formalism for TEGR has been presented in several references \cc{maluf1999general,blagojevic2000hamiltonian,maluf2001hamiltonian,da2010hamiltonian,okolow2014adm,ferraro2016hamiltonian}, however, we will focus on the approach introduced in \cc{ferraro2016hamiltonian}, and apply the main mathematical tools developed on it to NGR. In Ref. \cc{ferraro2016hamiltonian} the Lagrangian of TEGR has been written explicitly in terms of the tetrad field instead of the common use of the torsion tensor, and it leads to the appearance of the constitutive tensor $\chi_{ab}^{\hphantom{ab}cedf}$ for TEGR, which is a rank-six tensor built only of invariant objects like the Kronecker delta $\delta^{a}_{b}$ and the Minkowski metric $\eta_{ab}$. This mathematical object can easily be generalized to account for the NGR Lagrangian, nonetheless, its physical meaning goes deep in the premetric approach of any physical theory. The premetric formulation idea consists in removing the gravitational potential within GR, that is, the metric tensor of spacetime, from the fundamental laws of classical electromagnetism and gravitation \cc{Hehl:2016glb,Itin:2016nxk,Itin:2018dru}.

The mathematical formalism introduced in \cc{ferraro2016hamiltonian} has proven to be very useful for the Hamiltonian formulation of TEGR, because it maps the problem of finding the primary constraints of the theory to solving a problem of linear algebra, which only consists in finding the null kernel of the Hessian matrix. Later, the Hessian can be inverted through the calculation of the pseudoinverse matrix, which allows to solve for (some) canonical velocities and allows to obtain the Hamiltonian of the theory. The NGR case is by far more complicated, as the three free parameters in the Lagrangian of the theory allows nine cases with different number of primary constraints. In this work we would like to apply the mathematical tools developed in \cc{ferraro2016hamiltonian} to the NGR case, identify all possible cases and the number of primary constraints, and set the stage for future investigation.

This note is organized as follows. We introduce the basics of the teleparallel formalism and the NGR Lagrangian in terms of the constitutive tensor in Sec.\ref{sec:ngr}. We derive the relation among velocities and momenta, determine the Hessian tensor, and define the mapping to a Hessian matrix in Sec. \ref{sec:mom}. We study the mathematical properties of the Hessian, determine the null eigenvalues and define all possible cases in Sec. \ref{sec:eig}. We end up with discussion and conclusions in Sec. \ref{sec:con}.

\section{New general relativity in the  premetric approach}
\label{sec:ngr}

We begin defining a space with absolute parallelism, which will consist in a 4-dimensional manifold, where each point is labeled by a set of 4 independent coordinates $x^\mu$. Each point of this space has a linearly independent vectorial base $e_a = e^{\mu}_a \partial_{\mu}$  and a dual base of 1-forms $E^{a} = E^a_{~\mu} dx^{\mu}$ defined in its tangent space. Their components satisfy the orthonormality condition \cite{Aldrovandi:2013wha}
\begin{equation}\label{gmunuhh}
  \eta_{ab} = g_{\mu\nu} e^\mu_{~a} e^\nu_{~b}
\end{equation}
from which the metric is retrieved
\begin{equation}
    g_{\mu\nu}= \eta_{ab} E^a_{~\mu} E^b_{~\nu}.
\end{equation}
We will denote the coordinate components by Greek indices $ \mu, \nu, \ldots = 0, \ldots , 3$, and  Lorentzian tangent space  by Latin indices  $a, b, \ldots , g, h = 0, \ldots ,3$. The Minkowski metric will be $\eta_{ab}=\text{diag}(1,-1,-1,-1)$. \\

Our manifold will be equipped with an affine connection that is flat, and hence allows the parallel transport of any vector throughout the manifold, that is, it defines absolute parallelism. A generic way of getting a connection with zero curvature is the so-called inertial spin connection $\omega^{a}_{\ b\mu}=-(\Lambda^{-1})^{c}_{b}\partial_{\mu}\Lambda^{a}_{c}$, so that the components of the connection are \cite{golovnev2018introduction}
\begin{equation}\label{wtznbckcnnctn}
\Gamma^{\alpha}_{\mu\nu}  = e_{a}^{\;\;\alpha}( \partial_{\mu} E^{a}_{\;\;\nu}+\omega^{a}_{\ \mu b} E^{b}_{\;\;\nu}) .
\end{equation}
It is known that teleparallel gravity can be consistently formulated in the Weitzenb\"{o}ck gauge, where the spin connection is set to zero, $\omega^{a}_{\ b\mu}=0$, therefore the components of the  connection \eqref{wtznbckcnnctn} become 
\begin{equation}
\label{wtznbckcnnctnwithoutspin}
\Gamma^{\alpha}_{\mu\nu}  = e_{a}^{\;\;\alpha} \partial_{\mu} E^{a}_{\;\;\nu}.
\end{equation}
With the connection \eqref{wtznbckcnnctnwithoutspin}, one can define the components of the torsion and contortion tensors, respectively, as:
\begin{align}
T^{\alpha}_{\;\;\mu\nu}& = \Gamma^{\alpha}_{\nu\mu}-\Gamma^{\alpha}_{\mu\nu} = e_{a}^{\;\;\alpha}\left(\partial_{\mu} E^{a}_{\;\;\nu}-\partial_{\nu} E^{a}_{\;\;\mu}\right)\label{tor}\;,\\
K^{\mu\nu}_{\;\;\;\;\alpha} & = -\frac{1}{2}\left(T^{\mu\nu}_{\;\;\;\;\alpha}-T^{\nu\mu}_{\;\;\;\;\alpha}-T_{\alpha}^{\;\;\mu\nu}\right)\label{cont}\; .
\end{align}
For facilitating the description of the Lagrangian and the equations of motion, one can define another tensor from the components of the torsion and contortion tensors, which is known as superpotential, as \cite{Maluf:2013gaa}
\begin{equation}
S_{\alpha}^{\;\;\mu\nu}=\frac{1}{2}\left( K_{\;\;\;\;\alpha}^{\mu\nu}+\delta^{\mu}_{\alpha}T^{\beta\nu}_{\;\;\;\;\beta}-\delta^{\nu}_{\alpha}T^{\beta\mu}_{\;\;\;\;\beta}\right)\label{s}\;.
\end{equation}
Given this, we define the torsion scalar for the teleparallel equivalent of general relativity as
\begin{equation}
\label{t1}
T= T^{\alpha}_{\;\;\mu\nu}S_{\alpha}^{\;\;\mu\nu},
\end{equation}
which can alternatively be written as a quadratic form in the torsion tensor in the following way  
\begin{equation}
\label{tscalar}
T = \frac{1}{4}  T^{\rho}_{\ \mu\nu} T_{\rho}^{\ \mu\nu} - \frac{1}{2} T^{\rho}_{\ \mu\nu} T^{\mu\nu}_{\ \ \rho} - T^{\rho}_{\ \mu\rho} T^{\sigma\mu}_{\ \ \sigma}.
\end{equation}
The torsion scalar forms part of the TEGR Lagrangian as $L=ET$ (where $E=\text{det}(E^{a}_{\mu})$), which gives equations of motion that are equivalent to Einstein's equations. The simplest linear modification to \eqref{tscalar} consists in generalizing the coefficients in front of the three terms quadratic in the torsion tensor. This defines a generalized torsion scalar 
\begin{equation}\label{gtt}
T = c_1 T^{\rho}_{\ \mu\nu} T_{\rho}^{\ \mu\nu} +c_2 T^{\rho}_{\ \mu\nu} T^{\nu\mu}_{\ \ \rho} + c_3 T^{\rho}_{\ \mu\rho} T^{\sigma\mu}_{\ \ \sigma}
\end{equation}
that defines the NGR Lagrangian $L=ET$. Considering that the torsion tensor can be rewritten as 
\begin{equation}
T^{\rho}_{\ \mu\nu} = \partial_{\sigma}E^{a}_{\lambda} e^{\rho}_{a} e^{\sigma}_{b} e^{\lambda}_{c} E^{d}_{\mu} E^{e}_{\nu}(\delta^{b}_{d} \delta^{c}_{e} - \delta^{b}_{e} \delta^{c}_{d} ),
\end{equation}
and after some tensorial algebra, it is possible to write the NGR Lagrangian in an alternative form
\begin{equation}
L = \dfrac{1}{2} E \partial_{\mu} E^{a}_{\nu} \partial_{\rho} E^{b}_{\lambda} e^{\mu}_{c} e^{\nu}_{e} e^{\rho}_{d} e^{\lambda}_{f} \chi_{ab}^{\ \ cedf}.
\label{LNGR}
\end{equation}
An analogue procedure has been performed in Ref. \cite{ferraro2016hamiltonian} but for TEGR. Now, the mathematical object
\begin{equation}
\chi_{ab}^{\ \ cedf} = 4 c_1 \eta_{ab} \eta^{c[d} \eta^{f]e} + 4c_2 \delta_{a}^{[d}\eta^{f][e}\delta^{c]}_{b} + 4 c_3 \delta_{a}^{[c} \eta^{e][d}\delta^{f]}_{b},
\end{equation}
corresponds to the constitutive tensor for NGR, which has arbitrary coefficients $c_i, i=1,\ldots,3$. 

If we relax the requirement that the quadratic combinations of the torsion tensor in \eqref{gtt} has to preserve parity, then we can add two additional  parity-violating terms \cite{Ortin:2004,jimenez2020non}
\begin{equation}
c_4 \epsilon^{\mu\nu\lambda\rho} T_{\mu\nu\sigma} T^{\lambda\rho\sigma} , \qquad c_5 \epsilon^{\mu\nu\lambda\rho} T^{\sigma}{}_{\sigma\mu} T_{\nu\lambda\rho},
\end{equation}
which are related by an integration by parts. These terms have the following contribution to the constitutive tensor
\begin{equation}
4 c_4 \eta_{ag} \eta^{h[d}\eta^{f][e} \epsilon^{c]ghb} + 4 c_5 \eta_{ag}\delta_b^{[d} \epsilon^{f]gce}.
\end{equation}
Although we will not consider these terms in our Hamiltonian analysis, we observe they are easy to work with, as they just consist in an extension of the constitutive tensor. They could be considered in future work, as there is some recent interest in gravity including  parity-violating terms \cite{Chatzistavrakidis:2020wum,Li:2020xjt}

The constitutive tensor is an important object in the premetric approach to physics \footnote{For an introduction and applications of the premetric approach, see \cite{Hehl:2016glb}}. A layperson's description of the premetric approach is the following: it consists in a formulation of a physical theory that prescinds from the metric tensor as the fundamental variable. The building blocks of the premetric approach are a conserved charge $J$ such that its exterior derivative  $dJ=0$. For it, there exists a 2-form excitation field $H$ and a 2-form field strength $F$ that satisfy the fundamental equations of motion
\begin{equation}
dH=J, \ \ \ dF=0.
\end{equation}
These equations have physical meaning only when combined with a constitutive relation $H=\chi[F]$, which is the only relation that has a dependence on the metric. The details of the premetric approach for the teleparallel equivalent of general relativity can be studied from \cite{Itin:2016nxk,Itin:2018dru}. We claim  that our Hamiltonian formalism is performed in the premetric approach because we will make extensive use of the constitutive tensor $\chi_{ab}^{\ \ cedf}$ in the classification of primary constraints. In future work, we expect to use the mathematical properties of the constitutive tensor in the  determination of the primary constraints and the Hamiltonian of some subcases of NGR. The use of the constitutive tensor in the Hamiltonian formalism has not been explored in the literature, except for the TEGR case in Ref. \cite{ferraro2016hamiltonian}. In this sense, our work offers a new perspective and consists in a novel approach.

\section{Canonical momenta}

\label{sec:mom}

We are interested in determining the number of primary constraints for all cases allowed by the free parameters in the NGR Lagrangian. For this, we apply the well-known Dirac-Bergmann algorithm for constrained Hamiltonian systems \cc{anderson1951constraints,dirac1964lectures,sundermeyer1982constrained,henneaux1994quantization,sundermeyer2014symmetries} to the NGR Lagrangian discussed in the previous section. We will provide the mathematical tools needed for the identification of primary constraints, however, in order to find secondary constraints, computing the time evolution of them, and its subsequent classification in first and second class constraints, we will need to compute the Hamiltonian. Since this is a long task we leave for future work, and for now focusing on explaining the mathematical procedure and identifying all possible cases and the number of primary constraints in each one of them.

The canonical momenta for NGR are obtained straightforwardly from the Lagrangian \eqref{LNGR}, when taking the functional derivative with respect to $\partial_0 E^{a}_{\mu}$. We get
\begin{equation}
\Pi^{\mu}_{a} = \dfrac{\delta L}{\delta(\partial_0 E^{a}_{\mu})} = E \partial_{\rho} E^{b}_{\lambda} e^{0}_c e^{\mu}_{e} e^{\rho}_{d} e^{\lambda}_{f} \chi_{ab}^{\ \ cedf}.
\label{momb}
\end{equation}
Notice that unlike previous work, we do not consider an explicit ADM decomposition on the tetrad field \footnote{Also for GR the ADM decomposition of the metric is not essential, see discussion in Ref.\cite{Kiriushcheva:2008sf}.}, and only a time foliation is assumed at the moment we define the momenta \eqref{momb}. On this expression, we can isolate the time derivative of the tetrad $\partial_0 E^{a}_{\mu}$, by splitting the $\rho$ index in time and spatial components, obtaining
\begin{equation}
\label{mom-ab}
\Pi^{\mu}_{a} E^{e}_{\mu} = E C_{ab}^{\hphantom{ab}ef} e^{\lambda}_{f}\partial_{0}E^{b}_{\lambda} + E \partial_{i}E^{b}_{\lambda} e^{0}_{c} e^{i}_{d} e^{\lambda}_{f} \chi_{ab}^{\hphantom{ab}cedf },
\end{equation}
where the object
\begin{equation}
C_{ab}^{\hphantom{ab}ef } = e^{0}_{c}e^{0}_{d} \chi_{ab}^{\hphantom{ab}cedf }
\end{equation}
is the tensorial analog of a Hessian matrix.

At this point it is convenient to regard the multi-index notation introduced in \cc{ferraro2016hamiltonian} in order to rewrite the tensorial object $C_{ab}^{\ \,\,\,\,ef}$ as a
symmetric $16\times 16$ matrix. For this, we adopt a notation where we take pairs of flat indices $a,b,...$ to define a multi-index $A=\left( {}\right)
_{\hphantom{a} e}^{a}$, with the help of the following indexation formulas 
\begin{equation}
A=4a+(e+1),\ \ \ \ B=4b+(f+1);  \label{indexation}
\end{equation} 
so $A=\left( {}\right)
_{\hphantom{a}e}^{a},B=\left( {}\right) _{\hphantom{b} f}^{b},...=1,...,16$. This implies that $C_{AB}$ is a symmetric matrix, that is
\begin{equation}
C_{AB} = C_{BA}.  
\label{Csym}
\end{equation}
The formula \eqref{indexation} can be inverted by taking $a=[A/4]$, so $e=A-4[A/4]-1$, where $[\ ]$ means the integer part.

The purpose of the indexation formulas is to rewrite the expression for the momenta \eqref{mom-ab} in a friendlier way for linear algebra analysis. For this, we redefine all terms so that they are only expressed by means of the super-indices $A$. We achieve this by defining
\begin{equation}
\begin{split}
\dot{E}^{B} & = e_{f}^{\lambda} \dot{E}_{\lambda}^{b}, \\
E_{0}^{B} & = e_{f}^{i} \partial_{i}E_{0}^{b} \\
\Pi_{A} & = \Pi_{a}^{\mu} E_{\mu}^{e} \\
P_{A} & = E \partial_{i} E_{k}^{b} e_{c}^{0} e_{d}^{i} e_{f}^{k} \chi_{ab}^{\hphantom{ab}cedf}.
\end{split}
\end{equation}
In this way, the expression for the canonical momenta acquires a very simple form
\begin{equation}
\label{momA}
\Pi_{A}-P_{A} = E C_{AB}(\dot{E}^{B}-E_{0}^{B}).
\end{equation}
As we will see later, the matrix $C_{AB}$ is not invertible, therefore  not all velocities $\dot{E}^{B}$ are solvable in terms of the momenta $\Pi_{A}$, which is to be  expected in a constrained theory as NGR. However, the number of constraints will depend on the mathematical properties of the Hessian matrix, which in turn heavily depends on the free parameters of the theory.

It will be useful to define a mathematical object that raises and lowers super-indices $A$ in the same way that the Minkowski metric $\eta_{ab}$ raises and lowers indices belonging to the tangent space. Such Minkowski matrix in the superspace is defined as
\begin{equation}
\eta_{AB} = \eta_{ab} \eta^{ef}.
\end{equation}
This implies that the object $C^{B}_{\hphantom{B}A} = \eta^{BC} C_{CA}$ will not be symmetric in the pair of superindices $AB$, after noticing that
\begin{equation}
\label{superminkws}
\eta_{AB}=\text{diag}(1,-1,-1,-1,-1,1,1,1,-1,1,1,1,-1,1,1,1).
\end{equation}

Generically the matrices $C_{AB}$ and $C^{B}_{\hphantom{B}A}$ have different eigenvalues, but they share their null eigenvalues, as expected. Following \cc{ferraro2016hamiltonian}, we will calculate the eigenvalues of $C^{B}_{\hphantom{B}A}$, since it naturally comes as an invariant object in the superspace delimited by the Minkowskian supermetric $\eta_{AB}$.

\section{Eigenvalues of the Hessian matrix $C^{B}{}_{A}$}

\label{sec:eig}

An estimation of the number of primary constraints of the theory is possible through the computation of the null eigenvalues of the matrix $C^{B}_{\hphantom{B}A}$. Also, they will be helpful for the calculation of the Hamiltonian, an issue that we will leave for future work. The eigenvalues of the most general Hessian matrix for NGR (when all parameters are free) have the following pattern
\begin{equation}
\begin{split}
& \lambda_1, \ldots, \lambda_4 = 0, \\
& \lambda_5, \ldots, \lambda_9 = (2 c_1  +c_2) \; g^{00}, \\
& \lambda_{10}, \ldots, \lambda_{12} = (2 c_1 -c_2) \; g^{00}, \\
& \lambda_{13} = (2 c_1 +c_2 + 3c_3) \; g^{00}, \\
& \lambda_{14}, \ldots, \lambda_{16} = (2 c_1 +c_2 + c_3) \; g^{00}.
\end{split}
\label{eigenv}
\end{equation}

In \eqref{eigenv} we observe four null eigenvalues $ \lambda_1, \ldots, \lambda_4 $, which signal the existence of the primary constraints $\Pi^{0}_{a} \approx 0$, which come from the non appearance of $\partial_{0}E^{a}_{0}$ in the Lagrangian. These constraints are associated with four components of the kernel of $C^{B}_{\hphantom{B}A}$ (and also, of $C_{AB}$). In this simple case such vectors can be parameterized as $v^{A} = v_{|g|e}^{\ \ a} = e^0_e \delta^a_g$, and then for any $c_i$ it is accomplished that $ v^{A}  C^{B}_{\hphantom{B}A} = 0$. This can be written as
\begin{equation}
 v_{|g|e}^{\ \ a} C^{b \ \ e}_{\ af} = e^0_e \delta^a_g e^0_c e^0_d \chi^{b \ c\hphantom{f}de}_{\ a\hphantom{c} f}=0,
\end{equation}
or also $v^{A} C_{AB} = 0$, this is
\begin{equation}
 e^0_e \delta^a_g C_{ab}^{\ \ ef} =  e^0_e \delta^a_g e^0_c e^0_d \chi_{ab}^{\ \ cedf} = 0,
\end{equation}
which is straightforward from the fact that $\chi^{b \ c\hphantom{f}de}_{\ a\hphantom{c} f}$ is antisymmetric in $d-e$ but multiplied by $e^0_d e^0_e$, or that   $\chi_{ab}^{\ \ cedf}$ is antisymmetric in $c-e$, but multiplied by $e^0_e e^0_c$ which is symmetric on those indices.

After these considerations we are left with  $12$ nonvanishing eigenvalues $\lambda_5,\ldots,\lambda_{16}$. They are classified as follows: five of them are $(2 c_1 +c_2) \; g^{00}$, three are equal to $(2 c_1 -c_2) \; g^{00}$, one eigenvalue is $(2 c_1  +c_2 + 3 c_3) \; g^{00}$ and the remaining three are equal to $(2 c_1 +c_2 + c_3) \; g^{00}$. In relation to Ref. \cc{blixt2019ahamiltonian}, we obtain that the coefficients $A_i$ defined there are equivalent, up to normalization factors, to the eigenvalues we have obtained within our method. Here  $i=\mathcal{V},\mathcal{A},\mathcal{S},\mathcal{T}$ stand for vector, antisymmetric, symmetric and trace-free symmetric decomposition, respectively. The proportionality of those coefficients and our eigenvalues goes as follows:
\begin{equation}
\begin{split}
& A_{\mathcal{S}} \propto \lambda_{\mathcal{S}} = 2 c_1  +c_2,\\
& A_{\mathcal{A}} \propto \lambda_{\mathcal{A}} = 2 c_1  -c_2, \\
& A_{\mathcal{T}} \propto \lambda_{\mathcal{T}} = 2c_1 +c_2 + 3c_3, \\
& A_{\mathcal{V}} \propto \lambda_{\mathcal{V}} =  2c_1  +c_2 + c_3.
\end{split}
\end{equation}
This indicates that the $\mathcal{V,A,S,T}$ decomposition is intimately related with the algebraic structure of the Hessian matrix in the formalism we are presenting. Therefore, it will be interesting to explore the equivalence between both formalisms in future work. 

More primary constraints appear from vanishing one or more sets of eigenvalues $\lambda_i$ previously found. This is achieved by taking the adequate choice for the parameters $c_i$ in the NGR Lagrangian. All possible choices are summarized in Table \ref{summNGR}. We have numbered cases from 1 to 9 by following the order exhibited in unnumbered table of \cc{blixt2019ahamiltonian}.

\begin{table}[ht]
        \label{summNGR}
    \centering
    \begin{tabular}{c|c|c|c|c|c}
       case & null eigenvalue(s) & N$^o$ of p.c. & $c_1$ & $c_2$ & $c_3$ \\
       \hline
       case 1 & $\lambda_{i}\neq 0$ & 0 p.c. & free & free & free\\
       case 2 & $\lambda_{\mathcal{V}}=0$ & 3 p.c. & free & $2c_1+c_3$ & free \\
       
       case 3 & $\lambda_{\mathcal{A}}=0$ & 3 p.c.  & free & $-2c_1$ & free \\
        case 4 & $\lambda_{\mathcal{S}}=0$ & 5 p.c. & free & $2c_1$ & free \\
       case 5 & $\lambda_{\mathcal{T}}=0$ & 1 p.c. & free & $2c_1+3c_3$ & free \\

       case 6 & $\lambda_{\mathcal{V}}=\lambda_{\mathcal{A}}=0$ & 6 p.c. & free & $-2c_1$ & $-4c_1$ \\
        case 7 & $\lambda_{\mathcal{A}}=\lambda_{\mathcal{S}}=0$ & 8 p.c. & $0$ & $0$ & free \\
       case 8 & $\lambda_{\mathcal{A}}=\lambda_{\mathcal{T}}=0$ & 4 p.c. & $c_2/2$ & free & $-2c_2/3$\\
      
       case 9 & $\lambda_{\mathcal{T}}=\lambda_{\mathcal{S}}=\lambda_{\mathcal{V}}=0$ & 9 p.c. & free & $2c_1$ & 0 \\
       
    \end{tabular}
    \caption{All possible combinations of vanishing $\lambda_{i}$ and corresponding cases. }
\end{table}

Notice that other combinations where three eigenvalues are null are possible. For instance, it is possible to simultaneously vanish $(\lambda_{\mathcal{T}},\lambda_{\mathcal{A}},\lambda_{\mathcal{S}})$, or $(\lambda_{\mathcal{S}},\lambda_{\mathcal{A}},\lambda_{\mathcal{V}})$, or $(\lambda_{\mathcal{T}},\lambda_{\mathcal{A}},\lambda_{\mathcal{V}})$. However, they give trivial cases on which all $c_i=0$. Therefore, case 8 is the only nontrivial case which allows to simultaneously vanishing three eigenvalues.

We remark that the TEGR case is obtained when we substitute $c_1= \frac{1}{4}, c_2=-\frac{1}{2}$ and $c_3=-1$. We observe that this combination of parameters falls in case 6, where we already know that there are six primary constraints associated with local Lorentz transformations. Another interesting case is the one-parameter teleparallel gravity, which corresponds to case 2 with $c_3=1/2$ \cite{Cheng:1988zg}.

In future work we plan to build the canonical Hamiltonian for each case through the method of pseudoinverses described in \cc{ferraro2016hamiltonian}. For this, it will be helpful to replace the values of the $c_i$ appearing in Table \ref{summNGR}, for each case, in every  eigenvalue as appearing in Eq.\eqref{eigenv}. The results of this replacement can be found in Table \ref{eigenvaluesNGR}. 

\begin{table}[ht]
    \centering
    \begin{tabular}{c|c|c|c|c} 
       case & $\lambda_{\mathcal{S}}$ & $\lambda_{\mathcal{A}}$ & $\lambda_{\mathcal{T}}$ & $\lambda_{\mathcal{\nu}}$\\
       \hline
       case 1 & $2c_1+c_2$  & $2c_1-c_2$ & $2c_1+c_2+3c_3$  & $2c_1+c_2+c_3$\\
       case 2 & $-c_3$  & $4c_1+c_3$ & $2c_3$  & $0$ \\
       case 3 &  $4c_1$  & $0$ & $4c_1+3c_3$ & $4c_1+c_3$  \\
       case 4 & $0$  & $4c_1$ & $3c_3$  & $c_3$  \\
       case 5 & $-3c_3$ & $4c_1+3c_3$ & $0$ & $-2c_3$ \\

       case 6 & $4c_1$  & $0$ & $-8c_1$  & $0$ \\
       case 7 & $0$  & $0$ & $3c_3$  & $c_3$ \\
       case 8 & $2c_2$  & $0$ & $0$  & $4c_2/3$\\
      
       case 9 & $0$  & $4c_1$ & $0$  & $0$ \\
      
    \end{tabular}
    \caption{Eigenvalues for all possible cases }
    \label{eigenvaluesNGR}
\end{table}

In Table \ref{eigenvaluesNGR} we observe that some eigenvalues coincide, but evaluated in different cases. For example, the value of $\lambda_{\mathcal{S} }$ for case 6 is equal to the value of $\lambda_{\mathcal{A} }$ for case 4, although the multiplicity is different. We will explore this curious coincidence and its repercussion at the Hamiltonian level in future work.\\ \\ 

\section{Conclusions}

\label{sec:con}

We have applied a novel Hamiltonian approach in order to determine the number of primary constraints in the Hamiltonian formulation of NGR, and we have prepared the mathematical formalism for the calculation of the full set of constraints and the calculation of the  Hamiltonian. Firstly, we define the NGR Lagrangian in a premetric-like approach, by defining the constitutive tensor $\chi_{ab}^{\ \ cedf}$ which possess three free parameters $c_i$, and derive the canonical momenta in this formalism. We also obtained the components of the constitutive tensor related with parity-violating terms, which add two additional parameters $c_4$ and $c_5$,  and expect to study their dynamics in the future.  Afterwards, we obtain the expression for the canonical momenta in our formalism in Eq.\eqref{mom-ab}, and an expression in super-index notation in Eq.\eqref{momA}. We observe the appearance of the Hessian matrix $C_{AB}$ in the relation among the canonical velocities and momenta. Since this object depends on the constitutive tensor, we can use its mathematical properties to map the task of finding the primary constraints of the theory, to a problem of linear algebra, by searching for the null eigenvalues and null kernel of the Hessian $C$. 

We have calculated the eigenvalues of the $16 \times 16$ matrix $C^{B}_{\hphantom{B}A}$, and obtained four null eigenvalues for arbitrary $c_i$. These are related with four primary constraints associated with the absence of $\dot{E}^a_0$ in the Lagrangian. The remaining 12 eigenvalues come in four groups with multiplicity 3, 1, 5 and 3. Each of these groups can be vanished separately or in groups of two or three in a consistent way. The result of all the possible combinations give rise to nine cases (including arbitrary $c_i$), which are summarized in Table \ref{summNGR}. All the nine possible cases have different number and type of primary constraints, for different, restricted values of the $c_i$. Our results are in agreement with previous work in Ref.\cc{blixt2019ahamiltonian}. Finally, we have replaced back on the algebraic expression for the eigenvalues the particular values of the $c_i$ on each case, which is summarized in Table \ref{eigenvaluesNGR}. This computation will be useful in future work for the calculation of the canonical Hamiltonian on each case, which we intend to achieve through the method of pseudoinverses described in \cite{ferraro2016hamiltonian}.

The computation of the constraint algebra for all nine cases is an extense labour and will be addressed in future work. However, a clear understanding of the Hamiltonian formalism of NGR will be extremely useful for studying modified teleparallel gravities that have been proposed in the recent years, and can help to find good candidate theories that have good behaviour and well defined degrees of freedom.

\section*{Acknowledgments}
The authors would like to thank Daniel Blixt,  Waleed El Hanafy, Alexey Golovnev, Manuel Hohmann, Tomi Koivisto and Gamal Nashed for helpful discussions. The authors are grateful to the organizers and participants of the Teleparallel Gravity Workshop 2020, where this work has been presented. M.J.G  was funded by FONDECYT-ANID postdoctoral grant 3190531.

\appendix
\section{The Hessian matrix for NGR}
For the most general NGR Lagrangian, we write explicitly the Hessian matrix
\begin{tiny}
\begin{equation*}
C^{A}_{\ B} = \\
 \left(
\begin{tabular}{cccccccccccccccc}
\smallskip
 $D$ $f_{00}$ & $D$ $d_{01}$ & $D$ $d_{02}$ & $D$ $d_{03}$ & 0 & $c_{3}$ $c_{23}$ & -$c_{3}$ $d_{12}$ & -$c_{3}$ $d_{13}$  \\ \\
    \smallskip
 $-D$ $d_{01}$ &-$D$ ${e_{0}^{0}}^2- 2 c_{1} c_{23}$ & 2 $c_{1}$ $d_{12}$ & 2 $c_{1}$ $d_{13}$ & $c_{2}$ $c_{23}$ & 0 & $c_2$
   $d_{02}$ &  $c_2$ $d_{03}$  \\ \\
    \smallskip
 -$D$ $d_{02}$ & 2 $c_1$ $d_{12}$ & $-D$ ${e_{0}^{0}}^2- 2 c_1 c_{13}$ & 2 $c_1$ $d_{23}$ & $-c_2$ $d_{12}$ & -$ c_3$
   $d_{02}$ & $$C$$ $d_{01}$ & 0   \\ \\
    \smallskip
 $-D$ $d_{03}$ & 2 $c_1$ $d_{13}$ & 2 $c_1$ $d_{23}$ &-$D$ ${e_{0}^{0}}^2- 2 c_1 c_{12}$ & $-c_2$ $d_{13}$ & -$ c_3$
   $d_{03}$ & 0 & $$C$$ $d_{01}$ \\ \\
    \smallskip
 0 & $c_2$ $c_{23}$ & $-c_2$ $d_{12}$ & $-c_2$ $d_{13}$ & $$D$$ ${e^{0}_{1}}^2$-2 $c_1$ $c_{23}$ & $D$ $d_{01}$ & -2 $c_1$ $d_{02}$ & -2 $c_1$
   $d_{03}$ \\ \\
    \smallskip
 $ c_3$ $c_{23}$ & 0 & $c_3$ $d_{02}$ & $c_3$ $d_{03}$ & -$D$ $d_{01}$ & -$D$ ${f_{23}}$  & -$D$ $d_{12}$ & -$D$ $d_{13}$ \\ \\
    \smallskip
 -$c_3$ $d_{12}$ & $-c_2$ $d_{02}$ & -$$C$$ $d_{01}$ & 0 & 2 $c_1$ $d_{02}$ & -$D$ $d_{12}$ &$D$ ${e^{0}_{1}}^2- 2 c_1 c_{03}$ & 2
   $c_1$ $d_{23}$  \\ \\
    \smallskip
 -$c_3$ $d_{13}$ & $-c_2$ $d_{03}$ & 0 & -$$C$$ $d_{01}$ & 2 $c_1$ $d_{03}$ & -$D$ $d_{13}$ & 2 $c_1$ $d_{23}$ & $D$ ${e^{0}_{1}}^2- 2 c_1 c_{02}$  \\ \\

    0 & $-c_2$ $d_{12}$ & $c_2$ $c_{13}$ & $-c_2$ $d_{23}$ & $$C$$ $d_{12}$ & $c_3$ $d_{02}$ & $-c_2$ $d_{01}$ & 0

 \\ \\
  -$c_3$ $d_{12}$ & -$$C$$ $d_{02}$ & $-c_2$ $d_{01}$ & 0 & $c_2$ $d_{02}$ & 0 & $-c_2$ $c_{03}$ & $c_2$ $d_{23}$ 
 
 \\ \\ 
 
  $ c_3$ $c_{13}$ & $ c_3$ $d_{01}$ & 0 & $ c_3$ $d_{03}$ & -$ c_3$ $d_{01}$ & $c_3$ $c_{03}$ & 0 & -$c_3$ $d_{13}$
 \\ \\
 
 -$c_3$ $d_{23}$ & 0 & $-c_2$ $d_{03}$ & -$$C$$ $d_{02}$ & 0 & -$c_3$ $d_{23}$ & $c_2$ $d_{13}$ & $$C$$ $d_{12}$ 
 \\ \\
 
   0 & $-c_2$ $d_{13}$ & $-c_2$ $d_{23}$ & $c_2$ $c_{12}$ & $$C$$ $d_{13}$ & $c_3$ $d_{03}$ & 0 & $-c_2$ $d_{01}$
 \\ \\
 -$c_3$ $d_{13}$ & -$$C$$ $d_{03}$ & 0 & $-c_2$ $d_{01}$ & $c_2$ $d_{03}$ & 0 & $c_2$ $d_{23}$ & $-c_2$ $c_{02}$ 
 
 \\ \\
 
  -$c_3$ $d_{23}$ & 0 & -$$C$$ $d_{03}$ & $-c_2$ $d_{02}$ & 0 & -$c_3$ $d_{23}$ & $$C$$ $d_{13}$ & $c_2$ $d_{12}$
 
 \\ \\
 $c_3$ $c_{12}$ & $c_3$ $d_{01}$ & $c_3$ $d_{02}$ & 0 & -$c_3$ $d_{01}$ & $c_3$ $c_{02}$ & -$c_3$ $d_{12}$ & 0 \\ \\
     
\end{tabular} \right. \\
\end{equation*}

\begin{equation*}
 \left.
\begin{tabular}{cccccccccccccccc}

0 & -$c_{3}$ $d_{12}$ & $c_{3}$
   $c_{13}$ & -$c_{3}$ $d_{23}$ & 0 & -$c_{3}$ $d_{13}$ & -$c_{3}$ $d_{23}$ & $c_3$ $c_{12}$\\ \\
 
 $-c_{2}$ $d_{12}$ & $$C$$ $d_{02}$ & -$c_3$ $d_{01}$ & 0 & $-c_{2}$ $d_{13}$ & $$C$$ $d_{03}$ & 0 & -$c_3$ $d_{01}$ &
 \\ \\ 
  $c_2$ $c_{13}$ & $c_2$ $d_{01}$ & 0 & $c_2$ $d_{03}$ & $-c_{2}$ $d_{23}$ & 0 & $$C$$ $d_{03}$ & -$c_3$ $d_{02}$&

\\ \\

 $-c_2$ $d_{23}$ & 0 & -$c_3$ $d_{03}$ & $$C$$ $d_{02}$ & $c_2$ $c_{12}$ & $c_2$ $d_{01}$ & $c_2$ $d_{02}$ & 0 \\ \\
 
$$C$$ $d_{12}$ & $-c_2$ $d_{02}$ & $c_3$ $d_{01}$ & 0 & $$C$$ $d_{13}$ & $-c_2$ $d_{03}$ & 0 & $c_3$ $d_{01}$

\\ \\

     -$c_3$ $d_{02}$ & 0 & $c_3$
   $c_{03}$ & -$c_3$ $d_{23}$ & -$c_3$ $d_{03}$ & 0 & -$c_3$ $d_{23}$ & $c_3$ $c_{02}$

 \\ \\
   $c_2$ $d_{01}$ & $-c_2$ $c_{03}$ & 0 & $c_2$ $d_{13}$ & 0 & $c_2$ $d_{23}$ & $$C$$ $d_{13}$ & -$c_3$ $d_{12}$
 \\ \\
    
     0 & $c_2$ $d_{23}$ & -$c_3$ $d_{13}$ & $$C$$ $d_{12}$ & $c_2$ $d_{01}$ & $-c_2$ $c_{02}$ & $c_2$ $d_{12}$ & 0 \\ \\ 
 
  $D {e^{0}_{2}}^2-2 c_1 c_{13}$
   & -2 $c_1$ $d_{01}$ & $D$ $d_{02}$ & -2 $c_1$ $d_{03}$ & $$C$$ $d_{23}$ & 0 & $-c_2$ $d_{03}$ & $c_3$ $d_{02}$ \\ \\
    \smallskip
    
 2 $c_1$ $d_{01}$ & $D e^{0 2}_{2} -2c_1 c_{03}$ & -$D$ $d_{12}$ & 2 $c_1$ $d_{13}$ & 0 & $$C$$ $d_{23}$ & $c_2$ $d_{13}$ & -$c_3$ $d_{12}$ \\ \\
    \smallskip
    
  -$D$ $d_{02}$ & -$D$
   $d_{12}$ & -$D$ ${f_{13}}$ & -$D$ $d_{23}$ & -$c_3$ $d_{03}$ & -$c_3$ $d_{13}$ & 0 & $c_3$ $c_{01}$ \\ \\
    \smallskip
    
  2 $c_1$ $d_{03}$ & 2 $c_1$
   $d_{13}$ & -$D$ $d_{23}$ & $D e_{02}^2 -2c_1 c_{01}$ & $c_2$ $d_{02}$ & $c_2$ $d_{12}$ & $-c_2$ $c_{01}$ & 0 \\ \\
    \smallskip
    
   $$C$$ $d_{23}$ & 0 & $c_3$
   $d_{03}$ & $-c_2$ $d_{02}$ & $D e^{0 2}_{3} -2 c_1 c_{12}$ & -2 $c_1$ $d_{01}$ & -2 $c_1$ $d_{02}$ & $D$ $d_{03}$ \\ \\
    \smallskip
    
  0 & $$C$$ $d_{23}$ & -$c_3$
   $d_{13}$ & $c_2$ $d_{12}$ & 2 $c_1$ $d_{01}$ & $D e_{03}^2 -2c_1 c_{02}$ & 2 $c_1$ $d_{12}$ & -$D$ $d_{13}$ \\ \\
    \smallskip
    
  $c_2$ $d_{03}$ & $c_2$
   $d_{13}$ & 0 & $-c_2$ $c_{01}$ & 2 $c_1$ $d_{02}$ & 2 $c_1$ $d_{12}$ & $D e_{03}^2 -2c_1 c_{01}$ & -$D$ $d_{23}$ \\ \\
    \smallskip
    
  -$c_3$ $d_{02}$ &
   -$c_3$ $d_{12}$ & $c_3$ $c_{01}$ & 0 & -$D$ $d_{03}$ & -$D$ $d_{13}$ & -$D$ $d_{23}$ & -$D$ $f_{12}$ 
\end{tabular} \right)
\end{equation*}
\end{tiny}
where
\begin{equation}
\begin{split}
&f_{00}=(e^{0}_1)^2+(e^{0}_2)^2+(e^{0}_3)^2, \ \ \
f_{12}=(e^{0}_0)^2-(e^{0}_1)^2-(e^{0}_2)^2,\\
&f_{13}=(e^{0}_0)^2-(e^{0}_1)^2-(e^{0}_3)^2, \ \ \
f_{23}=(e^{0}_0)^2-(e^{0}_2)^2-(e^{0}_3)^2,\\
& c_{01}=(e^{0}_1)^2-(e^{0}_0)^2 , \ \ \
c_{02}=(e^{0}_2)^2-(e^{0}_0)^2, \\
& c_{03}=(e^{0}_3)^2-(e^{0}_0)^2 , \ \ \ c_{12}=(e^{0}_1)^2+(e^{0}_2)^2 , \\
& c_{13}=(e^{0}_1)^2+(e^{0}_3)^2 , \ \ \ c_{23}=(e^{0}_2)^2+(e^{0}_3)^2 , \\
& d_{01}=e^0_0 e^0_1 , \ \ \ d_{02}=e^0_0 e^0_2, \ \ \ d_{03}=e^0_0 e^0_3, \\
& d_{12}=e^0_1 e^0_2 , \ \ \ d_{13}=e^0_1 e^0_3 , \ \ \ d_{23}=e^0_2 e^0_3,\\
&D=-2c_1-c_2+c_3, \ \ \ 
C=-c_2+c_3.\\
\end{split}
\end{equation}

It is worth noticing that the $C^{A}_{\ B}$ matrix is not symmetric because we have raised one multi-index with the super-Minkowski metric $\eta^{AB}$, which has a pseudo Riemannian signature by blocks, as shown in Eq. \eqref{superminkws}.


\end{document}